\documentclass{article}

\usepackage{subcaption}
\usepackage{amsmath}
\usepackage{amsfonts}
\usepackage{amssymb}
\usepackage{mathtools}
\usepackage{graphicx}
\usepackage{textcomp}
\begin{document}

\huge{Detection of Aflatoxin M1 by Fiber Cavity Attenuated Phase Shift Spectroscopy}\\

\normalsize{M. Daniyal Ghauri$^{1}$, Syed Zajif Hussain$^{2}$, Ubaid Ullah$^{1}$, Rana M. Armaghan Ayaz$^{3}$, Rahman Shah Zaib Saleem$^{2}$, Alper Kiraz$^{3}$, and M. Imran Cheema $^{1,*}$} \\

$^1$Department of Electrical Engineering, Syed Babar Ali School of Science and Engineering, Lahore University of Management Sciences, Lahore 54792 Pakistan\\
$^2$Department of Chemistry and Chemical Engineering, Syed Babar Ali School of Science and Engineering, Lahore University of Management Sciences, Lahore 54792, Pakistan \\
$^3$Department of Physics, Ko\c{c} University, 34450 Sariyer, Istanbul, Turkey\\

$^*$imran.cheema@lums.edu.pk\\ 



\begin{abstract}
Aflatoxin M1 (AFM1) is a carcinogenic compound commonly found in milk in excess of the WHO permissible limit, especially in developing countries. Currently, state-of-the-art tests for detecting AFM1 in milk include chromatographic systems and enzyme-linked-immunosorbent assays. Although these tests provide fair accuracy and sensitivity however, they require trained laboratory personnel, expensive infrastructure, and many hours for producing final results. Optical sensors leveraging spectroscopy have a tremendous potential of providing an accurate, real time, and specialists-free AFM1 detector. Despite this, AFM1 sensing demonstrations using optical spectroscopy are still immature. Here, we demonstrate an optical sensor that employs the principle of cavity attenuated phase shift spectroscopy in optical fiber cavities for rapid AFM1 detection in aqueous solutions at 1550 nm. The sensor constitutes a cavity built by two fiber Bragg gratings. We splice a tapered fiber of $<$ 10 $\mu$m waist inside the cavity as a sensing head. For ensuring specific binding of AFM1 in a solution, the tapered fiber is functionalized with DNA aptamers followed by validation of the conjugation via FTIR, TGA, and EDX analyses. We then detect AFM1 in a solution by measuring the phase shift between a sinusoidally modulated laser input and the sensor output at resonant frequencies of the cavity. Our results show that the sensor has the detection limit of 20 ng/L (20 ppt) which is well below both the US and the European safety regulations. We anticipate that the present work will lead towards a rapid and accurate AFM1 sensor, especially for low-resource settings.
\end{abstract}

\section{Introduction}
\label{sec:introduction}
The feed of animals is often contaminated by a specific fungus which results in contamination of the produced milk with aflatoxin M1 \cite{Creppy_2002}. Aflatoxin M1 (AFM1) is a carcinogenic compound, which, if exceeded to a certain level, can cause serious health concerns to people \cite{WHO_AFM1_2017}. Given the hazardous nature of the contamination, many countries in the world have established guidelines for maximum permissible levels of AFM1 in milk. For example, the United States and European regulatory bodies limit the amount of AFM1 to 500 ng/L and 50 ng/L, respectively \cite{Creppy_2002}.

Currently, well-established methods of effective detection of AFM1 include thin-layer chromatography, high performance liquid chromatography, and immunoassays such as enzyme-linked immunosorbent assay \cite{yao2015developments}. The  frequent adaptation of these techniques, especially in the developing countries, is usually hindered due to the need for significant maintenance and operator expertise, extensive sample pre-treatments, time-consuming analyses, and high costs of infrastructures. A variety of optical sensors based upon surface plasmon resonance (SPR), interferometry, cavities, and fluorescence can overcome previously mentioned limitations. However, so far in the literature very few optical sensors have been reported for AFM1 detection.

In one of the promising approaches, researchers utilized long range SPR in conjunction with fluorophore-labeled AFM1 molecules for demonstrating a highly sensitive and rapid sensor \cite{wang2009long}. In another compact fluorometric system, researchers were able to provide preliminary screening of milk for AFM1 by using LED, detector, and associated optics \cite{cucci2007portable}. However, all the fluorescence-based techniques involve an arduous labeling process and locations of the fluorophores cannot be accurately controlled. This introduces uncertainty in the detected signal and creates significant challenges to produce repeatable and accurate results \cite{mahfuz2018general, hoa2007towards}. Moreover, apart from sensors based upon optical cavities, all others are generally single pass, i.e., an optical wave interacts with a sample only once. Whereas in a cavity based sensor, optical waves sample an analyte numerous times and thus highly accurate, sensitive, and rapid detection scheme can be realized. Motivated by this, researchers demonstrated a sensor based on silicon oxynitride ring resonators for AFM1 detection \cite{guider2015design}. They demonstrated 4.1 $\mu$g/L AFM1 detection in a DMSO solution. Their technique was based upon the frequency domain, i.e., they looked at the shift in a cavity resonant peak as a function of an AFM1 binding event. The use of frequency domain approach offers low noise immunity as compared to the time domain approach in optical cavities \cite{imran_2012}. The development of ring resonator based sensing platforms also has challenges in terms of accurately fabricating the ring and coupling waveguide with a precise gap. In order to overcome these drawbacks, recently researchers  demonstrated an Si$_3$N$_4$ asymmetric Mach-Zehnder interferometer (aMZI) for AFM1 detection \cite{chalyan2019afm1}. Although the aMZI sensor has the required limit of detection however, fabricating such a device is extremely challenging in low-resource settings where microfabrication facilities are generally not available. Motivated from this, we demonstrate an optical sensor for AFM1 detection that not only has the desired limit of detection but may also be easily developed and used in low-resource settings.

Cavity attenuated phase shift spectroscopy (CAPS) was first developed by Herbelin et al. in early 1980s for determining reflectivity of mirrors in free space cavities\cite{herbelin1981development,scherer1997cavity}. In CAPS, light from a CW laser source is amplitude modulated with angular frequency $\omega_m$ and is injected into a cavity. With respect to the input light, the output light from the cavity has a phase shift, $\phi$, that is related to the ring down time, $\tau$, of the cavity as
\begin{equation}
\tan \phi=-\omega_m \tau,
\end{equation}\label{eq:CAPS}
where the minus sign indicates that the cavity output lags the input. The constant monitoring of $\phi$ then gives continuous information about $\tau$ and consequently information about any attenuation events inside the cavity. CAPS and the one of its well-established variant, cavity ring down spectroscopy (CRDS), have been successfully applied in gaseous phase for various applications using free space cavities \cite{Berden_2000, Engeln1996,Kebabian_2005} .

For liquid phase applications, free space cavities pose a variety of challenges including instability of the cavity and degradation of mirrors due to the presence of liquid inside the cavity \cite{vallance2014cavity, thompson2017cavity}. These challenges have been overcome by utilizing microcavities \cite{barnes2013phase,imran_2012} and fiber based cavities and loops \cite{tong2004phase} for various liquid phase applications. Recently researchers applied CRDS in fiber loops for detecting aflatoxin B1 in a solution \cite{Chen_2019}. They did not apply any surface functionalization protocol to the sensing head and demonstrated the detection limit of 16 $\mu$g/L. CRDS is also inherently slow due to the requirement of the nonlinear curve fitting to extract the decay rate whereas in CAPS one can monitor sensing events continuously.

CAPS and any of its variants have never been utilized for detection of AFM1 either in free space or fiber based structures. Moreover, in all laser based fiber platforms utilizing CAPS, generally the Pound-Drever-Hall (PDH) technique is required for parking the laser for continuous monitoring of the phase shift. Here, we demonstrate an optical fiber sensor that not only utilizes the principle of CAPS for the first time to detect AFM1 but also involves a simpler experimental setup that does not require the use of the PDH technique. We show that our sensor has the detection limit of 20 ng/L (20 ppt) which is lower than the limits imposed by US and European authorities.

The organization of paper is such that, Section \ref{sec:Exp_proc} gives detailed experimental procedures which include  surface functionalization protocol and the sensor architecture. It is followed by Section \ref{sec:resul} which discusses results of the demonstrated sensor. Finally, Section \ref{sec:conc} concludes the paper along with some suggestions for future work.

\section{Experimental Procedures}\label{sec:Exp_proc}
In this section, we discuss various steps involved in developing the AFM1 sensor.

\subsubsection{Sensing head and its surface functionalization}\label{ssec:sen_set}

For developing the sensing head, we taper an optical fiber to expose its core for interaction of the optical wave with an analyte. The tapered fiber is fabricated by pulling an SMF-28 fiber with computer controlled motorized stages, while simultaneously heating the fiber with a butane flame. We fix the fabricated tapered fiber on a U-shaped piece of copper clad printed circuit board (PCB) using the nichrome soldering and then safely remove it from the tapering setup.  The use of alloy to fix the tapered fiber is critical as generally glues or epoxies malfunction with the application of abrasive chemicals like piranha during the functionalization steps. We adapt the the surface functionalization protocol reported in Ref. \cite{romain_2015} for our sensing head. The Ref. \cite{romain_2015} uses DNA aptamers for functionlizing SiN microring resonators, whereas we report the application of aptamers for conjugating silica tapered fibers.

In the present work, we first anneal the U-shaped copper PCB containing the silica tapered optical fiber at 1050$^{\circ}$C for 1.5 hours for removing any organic contaminants from the tapered fiber. For hydroxylation, we orient the U-shaped PCB in such a way that the fiber is dipped into a curved petri dish containing  the piranha solution. We then reperat the same procedure for silanization of the fiber with 0.01\%, V/V solution of 3-glycidoxypropyl methyldiethoxy silane prepared in anhydrous-toluene. The petri dish is then heated at 60$^{\circ}$C for a period of 10 min. The fiber is then immersed into 100 $\mu$M solution of a DNA aptamer having one end terminated with the amino group, 5AmMC6~/GTTGGGCACGTGTTGTCTCTCTGTGTCTCGTGCCCTTCGCTAGGCCCACA,  maintained in a phosphate buffer having pH of 8. All chemicals used in the functionlaization process are purchased from Sigma Aldritch Inc. USA except the aptamer which is obtained from IDT integrated DNA technologies Inc. USA.

\begin{figure}
\centering
\begin{subfigure}{.5\textwidth}
  \centering
  \includegraphics[width=\textwidth]{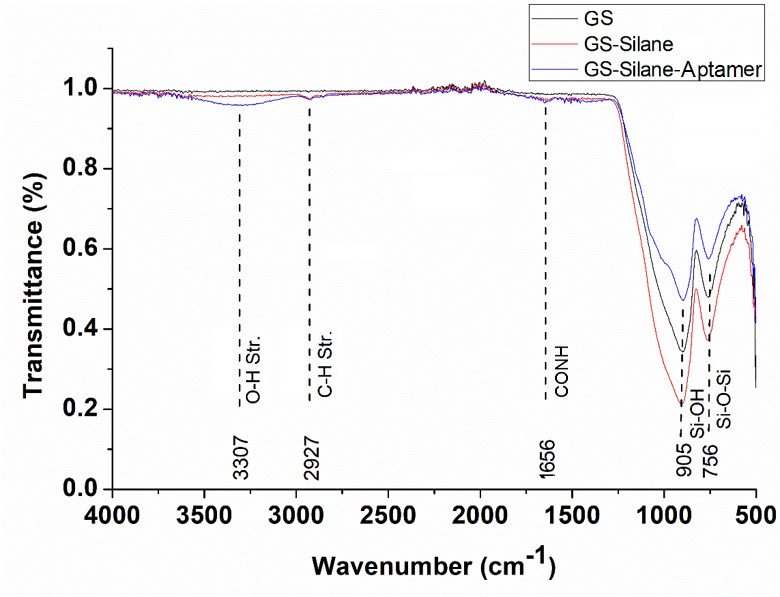}
  \caption{}
  \label{fig:FTIR}
\end{subfigure}%
\begin{subfigure}{.5\textwidth}
  \centering
  \includegraphics[width=\textwidth]{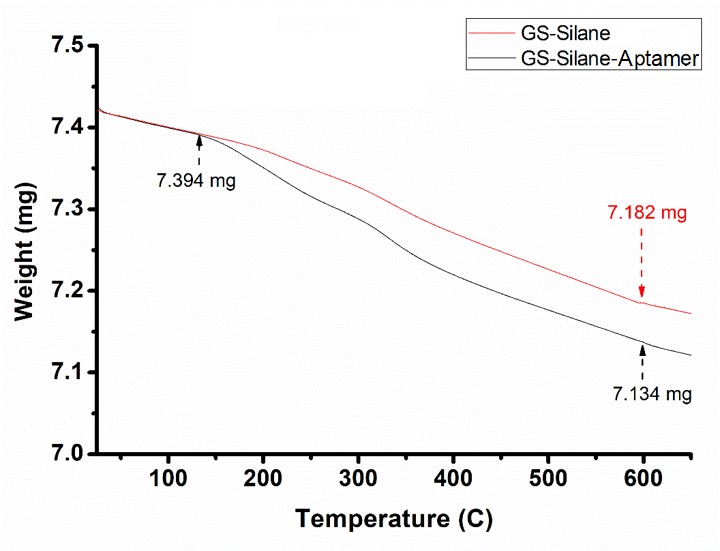}
  \caption{}
  \label{fig:TGA}
\end{subfigure}
\caption{(a) FTIR results showing silanization and conjugation of DNA apatmers over the surface of a glass substrate (GS). (b) Thermogravimetric analysis presenting weight loss of silanes and silanes + DNA-Aptamer conjugated separately over the surface of an SMF-28 optical fiber}
\label{fig:suf_fucs}
\end{figure}

The initial implementation of the protocol is done on a glass slide since its surface properties are very similar to a silica fiber. The FTIR results presented in Fig. \ref{fig:FTIR} show that the silanization followed by the conjugation of DNA apatmers are successfully achieved on the glass substrate. The characteristic peaks appearing in the substrate's FTIR spectrum at 756 and 905 cm$^{-1}$ correspond respectively to its Si-OH and Si-O-Si functionalities. After the silanization process, the intensities of these peaks rise indicating an increase in the number of these functionalities over the substrate surface. In addition to this, the peaks appearing between 2840 and 3000 cm$^{-1}$ also narrate the presence of C-H bonds signaling the successful silanization. Besides this, the conjugation of DNA aptamer on silanized glass slide is verified by the suppression of peaks at 756 and 905 cm$^{-1}$, and the respective appearance of O-H stretching and amide bands at 3307 and 1656 cm$^{-1}$.

In the second phase, the protocol is implemented on a non-tapered SMF-28 fiber. The thermogravimetric analysis (TGA) for the processed fiber in Fig. \ref{fig:TGA} indicates that the developed surface chemistry protocol is working on the fiber. This is confirmed by noting the loss of weight in the fiber conjugated with the DNA aptamer as compared to the silanized fiber. The reduction in weight is due to the burning of organic contents added by the DNA aptamer.

\begin{figure*}
\centering
\captionsetup{justification=centering}
\includegraphics[width=\textwidth]{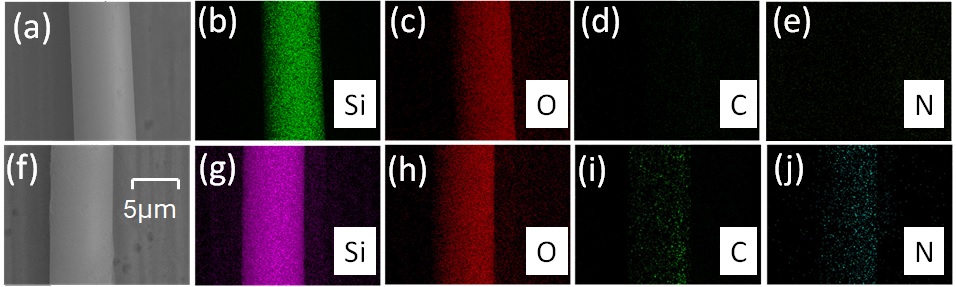}
\caption{(a-e) and (f-j) are EDX mapping analysis of the non-functionalized and functionalized tapered fibers, respectively. The presence of silicon (Si) and oxygen (O) shows silica fibers in both of the cases. Carbon (C) and nitrogen (N) are present only in the functionalized fiber due to silanes and DNA aptamers. The scale is 5$\mu$m for all images.}
\label{fig:edx}
\end{figure*}

The protocol is implemented on tapered fibers in the third phase of the surface functionliaztion. Here, we utilize the energy-dispersive X-ray spectroscopy (EDX) for comparing the elemental compositions of non-functionalized and functionalized tapered optical fibers as shown Fig. \ref{fig:edx}. The presence of extra signals of carbon and nitrogen elements in the EDX mapping of the functionalized tapered fiber as compared to the non-functionalized one confirms successful conjugation process. 

\subsubsection{Experimental setup} \label{ssec:meaur_anal}
\begin{figure*}
\begin{center}
\centering
\captionsetup{justification=centering}
\includegraphics[width=\textwidth]{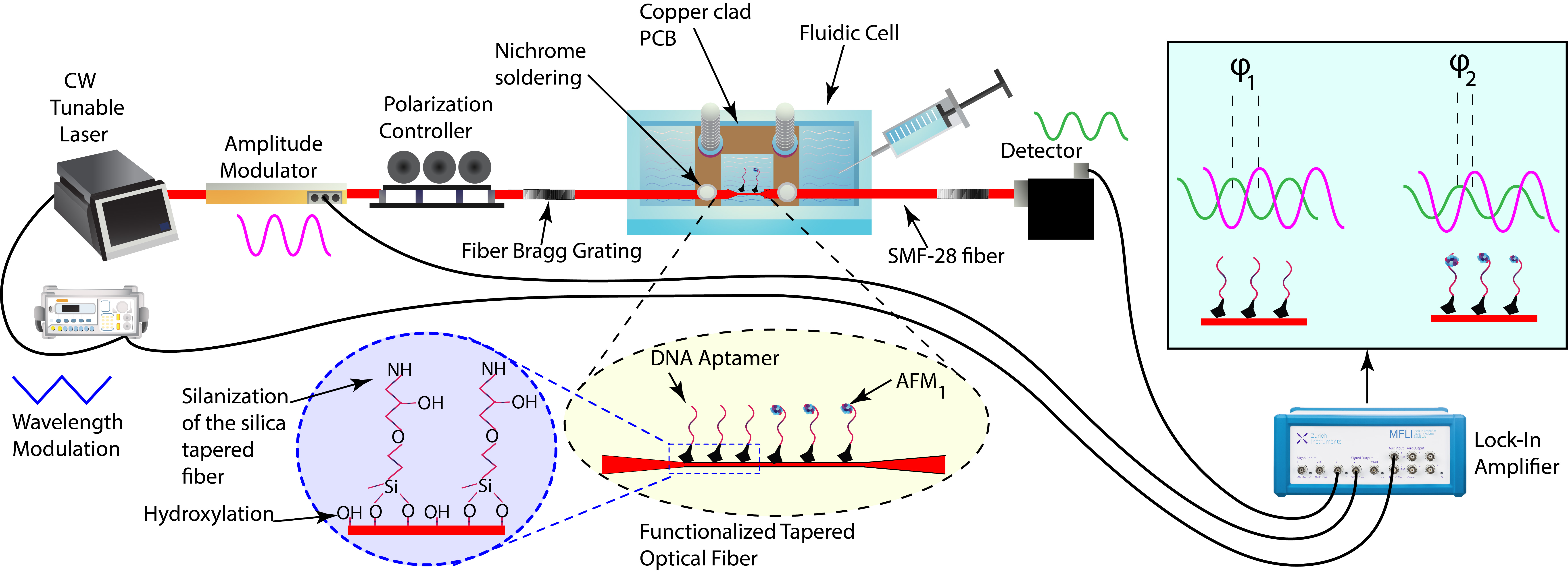}
\caption{Schematics (not drawn to scale) for the sensor for detecting AFM1 using the cavity attenuated phase spectroscopy (CAPS). The phase shifts and modulations depths are exaggerated for illustrating the CAPS measurement where $|\phi_1|>|\phi_2|$. }\label{fig:exp_set}
\end{center}
\end{figure*}

The schematics of our sensor for conducting CAPS measurements with fiber cavities is shown in Fig. \ref{fig:exp_set}.
We first build an optical fiber cavity by using fiber Bragg gratings (FBGs) at the two ends of an SMF-28 fiber. The two FBGs are purchased from Oeland Inc. Canada with parameters of FBG$_1$: 1549.99 nm, 91.49\%, 0.28 nm  and FBG$_2$: 1550.02 nm, 99.44\%, 0.23 nm. The first, second, and third parameters refer to the central wavelength, reflectivity, and full width half max bandwidth, respectively. The functionlized tapered fiber (waist diameter $\approx$ 8.5 $\mu$m) is then spliced between the two FGBs which results in the overall cavity length of 32 cm. We then package the U-shaped piece containing the tapered fiber portion into a 3D-printed fluidic cell for efficient delivery of a sample to the tapered portion.

The sensor source is a 1550nm continuous wave laser diode that is purchased from Eblana Photonics (EP1550-0-NLW-B26-100FM) and operated with Thorlabs CLD1015 controller. For initiating the CAPS measurements, we sinusoidally modulate (4 MHz, 1 V$_{pp}$) the laser using an external Mach-Zehnder modulator (Sumitomo Osaka T.MZH1.5-10PD-ADC). For scanning the cavity resonances, the laser is current tuned continuously using a triangular wave (5 mHz, 100 mVpp). The simultaneous current and amplitude modulations allow us to conduct CAPS measurements at resonant peaks without needing the PDH technique \cite{imran_2012}. The laser wavelength is tuned slow enough to allow a proper build-up of the cavity for conducting stable measurements. The modulated laser output is injected into the cavity and the same signal is used as a reference input signal to the lock-in amplifier (Zurich Instruments MFLI). An InGaAs photodetector (Thorlabs DET08CFC/M) is used for monitoring the cavity output which in turn is connected to the signal port of the lock-in amplifier.

The AFM1 binding events at the tapered fiber induce losses in the cavity; therefore, the phase change is recorded based upon CAPS. In traditional CAPS systems, one is required to use a source wavelength that matches the maximum absorption wavelength of an analyte to be probed. In our case, we are not restricted to use the absorption wavelength of the AFM1 as losses are mainly introduced due to the evanescent field of the tapered fiber. As a binding event occurs more evanescent field gets outside of the tapered fiber, hence losses are introduced in the cavity.

\section{Results}\label{sec:resul}

\begin{figure}
\centering
\captionsetup{justification=centering}
\includegraphics[width=0.8\textwidth]{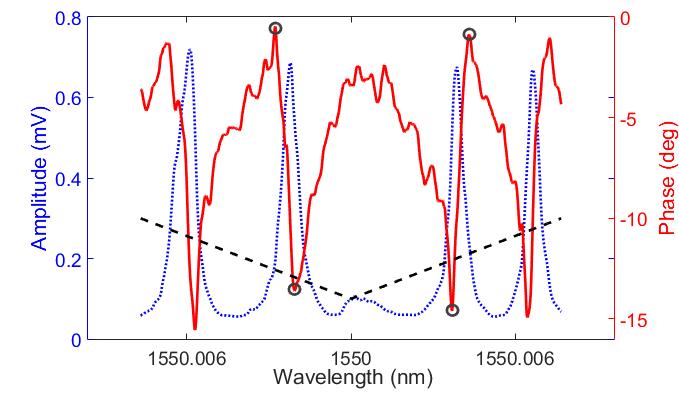}
\caption{Representative CAPS data. The black dashed curve represents a part of the triangular wave used for the wavelength modulation. The left and right hand sides of the 1550 nm point show downward and upward wavelength scans, respectively.  The blue dotted curve shows resonance peaks. The red solid curve shows phase measurements using our wavelength scanned CAPS technique.}
\label{fig:CAPS}
\end{figure}

A representative CAPS measurement for an FBG based cavity is shown in Fig. \ref{fig:CAPS}.
We observe that, in comparison to the position of a resonance peak, the absolute phase maximum is red-shifted and blue-shifted during the decreasing and increasing wavelength scans, respectively. This behavior can be explained by solving coupled mode equations for a transmission cavity \cite{Haus_1984} with a sinusoidally modulated laser input \cite{Cheema_2015}. We obtain the following solutions for the magnitude and phase of the output cavity field 
\begin{equation}
|E_{op}|=\dfrac{c \tau \sqrt{\left(1-R_1\right)\left(1-R_2\right)}}{2 n l\sqrt{1+\tau^2\left(\omega-\omega_o+\omega_m \right)^2}}|E_{in}|,
\label{eq:Mag_W_CAPS}
\end{equation}

\begin{equation}
\tan \phi=-\left(\omega-\omega_o+\omega_m \right) \tau,
\label{eq:phase_W_CAPS}
\end{equation}
where, $R_1$ and $R_2$ are reflectivities of the two FBGs, $c$ is the speed of light, $n$ is the SMF-28 fiber effective index, $l$ is the cavity length, $\omega_o$ is a resonant frequency of the cavity, and $\omega$ is the laser scan frequency. At a resonant frequency, $\omega=\omega_o$, Eqs. (\ref{eq:Mag_W_CAPS}) and (\ref{eq:phase_W_CAPS}) reduce to the classical CAPS equations. It is evident from Eq. (\ref{eq:Mag_W_CAPS}) that $|E_{op}|$ is always maximum whenever $\omega=\omega_o-\omega_m \approx \omega_o$. In the case of Eq. (\ref{eq:phase_W_CAPS}), it is clear that $|\tan \phi|$ is maximum whenever $\omega>\left(\omega_o-\omega_m \approx \omega_o\right)$ within a cavity resonance peak. This results in red-shift and blue-shift of a $|\phi_{max}|$, with respect to a resonance amplitude peak, for the upward and downward frequency scans, respectively. 

\begin{figure}
\centering
\begin{subfigure}{.5\textwidth}
  \centering
  \includegraphics[width=\textwidth]{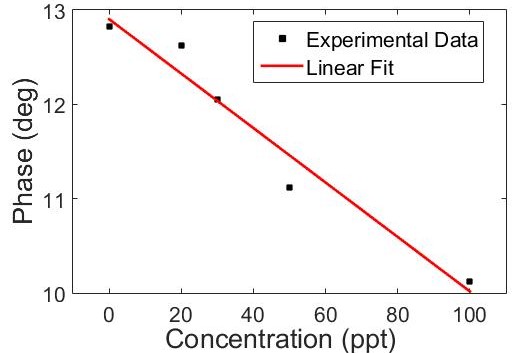}
  \caption{}
  \label{fig:AF_results}
\end{subfigure}%
\begin{subfigure}{.5\textwidth}
  \centering
  \includegraphics[width=\textwidth]{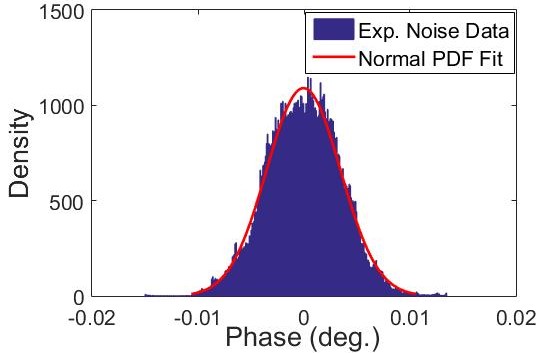}
  \caption{}
  \label{fig:noise}
\end{subfigure}
\caption{(a) Experimental results for detecting AFM1 using CAPS. (b) Noise data of the CAPS sensing setup}
\label{fig:exp_res}
\end{figure}

The AFM1 solution, derived from the aspergillus flavus, is purchased from Sigma-Aldrich Inc. USA. We then dilute it in DI water for preparing its various concentrations. We start our measurements with DI water followed by a known AFM1 concentration solution. We inject liquid samples into the fluidic cell by using a common syringe. After recording CAPS measurements with each AFM1 sample, we rinse the fluidic cell along with the tapered fiber first with a glycerine based buffer solution and then afterwards with DI water. This rinsing step removes any AFM1 binding event and consequently the functionalized tapered fiber is recycled for a new measurement. The phase shift data collected from the lock-in amplifier is analyzed in MATLAB. We subtract the maximum and minimum phase of each phase peak (representative data shown by circles in Fig. \ref{fig:CAPS}) and then take the mean of all the differences which gives one phase measurement for an injected sample. The measured phase due to CAPS for various concentrations of AFM1 solutions is shown in Fig. \ref{fig:AF_results}.

From these results we can see that our demonstrated CAPS sensor has the sensitivity and detection limit of 0.029$^o$/ppt and 20 ppt, respectively.  The noise in the sensor is measured by taking the laser out of the cavity resonance and continuously monitor the phase data coming from the sensor. The noise histogram data along with the normal probability distribution fit is shown in Fig. \ref{fig:noise}. The fit shows that one standard deviation, i.e., noise is 0.003 $^{\circ}$.

The concentration in an unknown sample can be estimated from the response curve accordingly to the phase observed for it. We also see that as the AFM1 concentration increases, the phase decreases. This trend is due to the increase in the cavity losses caused by the AFM1 binding events. We also try to measure similar AFM1 concentrations with a non-functionalized tapered fiber and find no phase changes. This indicates that functionalizing the sensing head is critical for detecting lower concentrations.

\section{Conclusion}\label{sec:conc}
In this work, we demonstrate the first application of CAPS to fiber cavities for detecting AFM1 in a solution. The experimental results obtained depict AFM1 detection as low as 20 ng/L which is well below the the safety limits laid down by both the US and the European regulatory authorities. In comparison to traditional and previous AFM1 sensors, the demonstrated sensor modality offers reduction in detection time and cost, minimum sample pre-treatment, decrease in system complexity, lower fabrication challenges, and increased potential for portability. There are various future directions for improving the present work. These include testing with actual milk samples, introducing temperature controlled environments for the FBGs and fluidic cell, improving and optimizing the tapering of fibers for enhancing the sensor sensitivity. We anticipate that the demonstrated sensing setup with alteration to the surface chemistry protocol will lead to an inexpensive, real-time and reliable optical sensor for numerous applications in food security, health, environmental, and agricultural sectors especially in low-resource settings.

\section*{Acknowledgment}

The described research activity is a part of the project funded by Pakistan Science Foundation (PSF) under grant no. PSF-TUBITAK/P-LUMS (3). We would also like to thank Hazem Asif for helping us in drawing Fig. \ref{fig:exp_set} of experimental schematics.

\section*{Disclosures}
The authors declare no conflicts of interest.

\bibliographystyle{IEEEtr}
\bibliography{references}

\end{document}